\documentclass{svmult}

\usepackage{latexsym}
\usepackage{amssymb}
\usepackage{epsfig}
\usepackage{fancyhdr}
\usepackage{float}
\usepackage[T1]{fontenc}
\usepackage[latin1]{inputenc}
\usepackage[bottom]{footmisc}
\usepackage{graphics}
\usepackage{graphicx}
\usepackage{multicol}
\usepackage{pslatex}
\usepackage{pifont}
\usepackage{supertabular}
\usepackage{t1enc}
\usepackage{units}
\usepackage{varioref}
\usepackage{wrapfig}
\usepackage{subfigure}
\usepackage{lscape}
\usepackage{slashbox}
\usepackage{longtable}
\usepackage{amsmath,eepic}
\usepackage{multirow}

\newcommand{\qq}[1]{{\lq}#1{\rq}}
\newcommand{\etal}{\textit{et al.}}

\newcommand{\absnum}{\ensuremath{N}}
\newcommand{\density}{\ensuremath{\rho}}
\newcommand{\inta}{\ensuremath{k}}
\newcommand{\liq}[1]{{#1}'}
\newcommand{\LJeps}{\ensuremath{\epsilon}}
\newcommand{\LJsig}{\ensuremath{\sigma}}
\newcommand{\mardyn}{\textit{ls1 mardyn}}
\newcommand{\mstwo}{\textsf{ms2}}
\newcommand{\radius}{\ensuremath{r}}
\newcommand{\radiusS}{\ensuremath{\radius_\mathrm{St}}}
\newcommand{\temperature}{\ensuremath{T}}
\newcommand{\vap}[1]{{#1}''}

\newcommand{\contactangle}{\ensuremath{\vartheta}}

\newcommand{\cpenergy}{\ensuremath{\FWenergy_0}}
\newcommand{\cutoff}{\ensuremath{\distance_\mathrm{c}}}
\newcommand{\degree}[1]{\ensuremath{{#1}^{\circ}}}

\newcommand{\distance}{\radius}
\newcommand{\FWenergy}{\ensuremath{\xi}}

\newcommand{\kboltz}{\ensuremath{k}}
\newcommand{\LJenergy}{\LJeps}

\newcommand{\LJsize}{\LJsig}

\newcommand{\LJTS}{TSLJ}
\newcommand{\moli}{\ensuremath{i}}
\newcommand{\molj}{\ensuremath{j}}

\newcommand{\potential}{\ensuremath{u}}

\newcommand{\potentialLJTS}{\ensuremath{\potential^\mathrm{\LJTS{}}}}

\newcommand{\temperaturecrit}{\ensuremath{\temperature_\mathrm{c}}}
\newcommand{\tension}{\ensuremath{\gamma}}
\newcommand{\tensionvl}{\ensuremath{\tension}}

\newcommand{\vapour}{vapor}
\newcommand{\Vapour}{Vapor}

\usepackage{makeidx}         
\usepackage{graphicx}        
\usepackage{multicol}        
\usepackage[bottom]{footmisc}

\makeindex             

\begin{document}

\title*{Molecular modeling of hydrogen bonding fluids: Vapor-liquid coexistence and interfacial properties}
\titlerunning{Molecular modeling of hydrogen bonding fluids}
\author{ Martin Horsch\inst{1} \and Martina Heitzig\inst{2}
         \and Thorsten Merker\inst{3} \and Thorsten Schnabel\inst{2}
	 \and Yow-Lin Huang \inst{1} \and Hans Hasse\inst{3} \and Jadran Vrabec\inst{1} }
\authorrunning{Horsch \etal}
\institute{
                 Lehrstuhl f\"ur Thermodynamik und Energietechnik (ThEt),
                 Universit\"at Paderborn, Warburger Str.\ 100,
                 33098 Paderborn, Germany\footnote{
                    Author to whom correspondence should be addressed:
                    Prof.\ Dr.-Ing.\ habil.\ J.\ Vrabec.
                    $\textnormal{E-mail}$: \texttt{jadran.vrabec@upb.de}.
                 }
	      \and
	         Institut f\"ur Technische Thermodynamik und Thermische Verfahrenstechnik (ITT),
		 Universit\"at Stuttgart, Pfaffenwaldring 9, 70569 Stuttgart, Germany
              \and
                 Lehrstuhl f\"ur Thermodynamik (LTD), Technische Universit\"at
                 Kaiserslautern, Erwin-Schr\"odinger-Str.\ 44,
                 67663 Kaiserslautern, Germany
          }

\maketitle

\section{Introduction}

A major challenge for molecular modeling consists in optimizing the unlike
interaction potentials. A broad study on fluid
mixtures \cite{SVH07} recently showed that
among the variety of combination rules that were proposed in the past,
none is clearly super\-ior. In many cases, all are suboptimal
when accurate predictions of properties like the mixture vapor pressure are needed.
The well known Lorentz-Berthelot rule performs quite well and can be used as a starting
point. If more accurate results are required, it is often advisable to adjust the 
dispersive interaction energy parameter which leads to very favorable
results \cite{SVH07, VSH05, UNRAL07, SVH08, VHH09}.

A similar approach should be followed for effective pair potentials acting
between fluid particles and the atoms of a solid wall. They can only be reliable
if fluid-wall contact effects are taken into account, e.g.,
by adjusting unlike parameters to contact angle measurements \cite{WWJHK03}.
Teletzke \etal{} \cite{TSD82} used continuum methods to examine
the dependence of wetting and dewetting transitions on characteristic size
and energy parameters of the fluid-wall dispersive interaction. MD simulation can
be applied for the same purpose, leading to a consistent molecular approach.

On the molecular level, the precise position of the vapor-liquid phase boundary
is defined by a cluster criterion. Many different criteria are known and
it is not im\-mediately obvious which of them leads to
the most accurate results \cite{WR07}.
In nanoscopic systems, minute absolute differences can lead to comparably large
relative deviations. Therefore, the viability of several criteria is compared
in the present study with the purpose of excluding errors
due to an inaccurate detection of the interface.

If the cohesion of the liquid phase is partly due to hydrogen bonds,
successful molecular models for pure fluids can often be developed on the basis of
an \textit{ab initio} study of the charge distribution as well as the equilibrium
position of the nuclei.
This sterically realistic approach, combined with adjusting the Lennard-Jones (LJ)
potential parameters to vapor-liquid equilibrium (VLE) data,
leads to empirical models that correctly reproduce and predict
thermophysical fluid properties over a wide range of conditions \cite{SCVLH07}.
The present work applies this approach to mixtures containing hydrogen bonding components.
Often potential parameters determined for one fluid carry over to
a derivative with different substituents, opening the possibility of 
creating generic molecular models. Such a model is presented for benzyl alcohol.

The following publications in peer-reviewed international journals
contribute to the present project:
\begin{itemize}
   \item{} Schnabel, T., Vrabec, J.\ \& Hasse, H.
           Molecular simulation study of hydrogen bonding mixtures and
           new molecular models for mono- and dimethylamine.
           \textit{Fluid Phase Equilib.}\ \textbf{263}: 144--159 (2008).
   \item{} Eckl, B., Vrabec, J.\ \& Hasse, H.
           An optimized molecular model for ammonia.
           \textit{Mol.\ Phys.}\ \textbf{106}: 1039--1046 (2008).
   \item{} Eckl, B., Vrabec, J.\ \& Hasse, H.
           Set of molecular models based on quantum mechanical ab initio
	   calculations and thermodynamic data.
           \textit{J.\ Phys.\ Chem.\ B} \textbf{112}: 12710--12721 (2008).
   \item{} Vrabec, J., Huang, Y.-L.\ \& Hasse, H.
           Molecular models for 267 binary mixtures validated by
	   vapor-liquid equilibria: a systematic approach.
           \textit{Fluid Phase Equilib.}\ \textbf{279}: 120--135 (2009).
   \item{} Huang, Y.-L., Vrabec, J.\ \& Hasse, H.
           Prediction of ternary vapor-liquid equilibria for
	   33 systems by molecular simulation.
           \textit{Fluid Phase Equilib.}\ \textbf{287}: 62--69 (2009).
   \item{} Horsch, M., Heitzig, M., Dan, C., Harting, J., Hasse, H.,
           Fischer, J.\ \& Vrabec, J. Contact angle dependence on the fluid wall dispersive energy.
	   In preparation.
\end{itemize}
It would exceed the scope of the present report to give a full exposition of these
articles. Instead, a few points are emphasized and arranged as follows:
Firstly, mixture properties are explored for binary systems containing hydrogen
bonding components. Secondly, vapor-liquid interface cluster criteria
and contact angles are discussed and remarks on computational details are given.
Finally, a sterically accurate generic model for benzyl alcohol is introduced
and evaluated.

\section{Fluid mixtures with hydrogen bonding components}

Vapor-liquid equilibria of 31 binary mixtures consisting of one hydrogen bonding
and one non-hydrogen bonding component were studied. All models are of the rigid
united-atom multi-center LJ type with superimposed electrostatic sites
in which hydrogen bonding is described by partial charges. The hydrogen bonding
components of the studied binary mixtures are: monomethylamine (MMA) and
dimethylamine (DMA), methanol, ethanol and formic acid.
The non-hydrogen bonding components are: neon, argon, krypton, xenon, methane,
oxygen, nitrogen, carbon dioxide, ethyne, ethene, ethane, propylene, carbon monoxide, diflouro\-dichloro\-methane (R12), tetra\-flouro\-methane (R14), diflouro\-chloro\-methane (R22),
diflouro\-methane (R32), 1,1,1,2-tetra\-flouro\-ethane (R134a) and 1,1-diflouro\-ethane (R152a).

\begin{figure}[h!]
\centering
\includegraphics[width=0.85\textwidth]{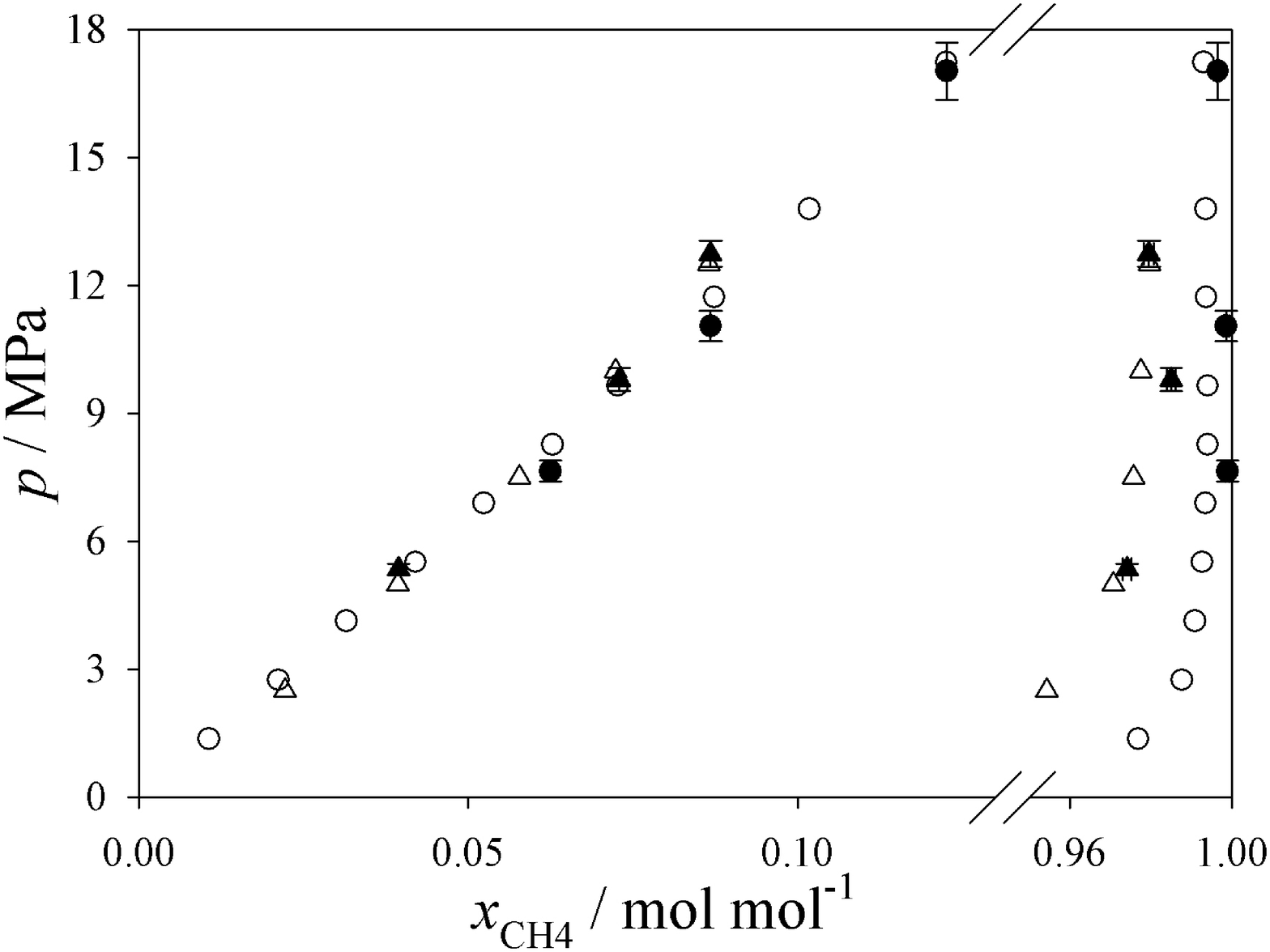}
\caption{
   Simulation data and vapor-liquid equilibria of methanol + methane:
   simulation data ($\bullet$) and experimental data ($\circ$) at 310 K \cite{HMJK87};
   simulation data ($\blacktriangle$) and experimental data ($\Delta$) at 338.15 K \cite{YSKL85}.
}
\label{figJVA}
\end{figure}

\begin{figure}[t!]
\centering
\includegraphics[width=0.85\textwidth]{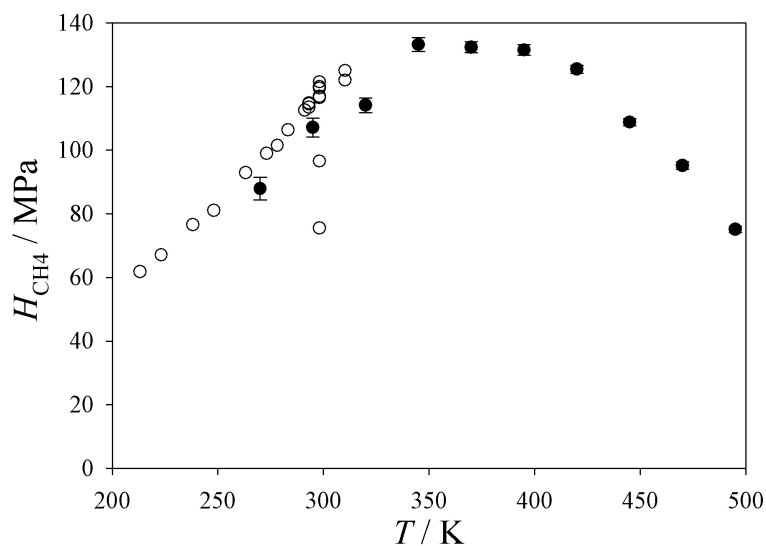}
\caption{
   Henry's law constant of methane in methanol:
   simulation data ($\bullet$) and experimental data
   \cite{Levi01, McDaniel11, BT58, BB60, LG60, SZI61, TK75, MRB79, Tkaczuk86, BBW93}
   ($\circ$).
}
\label{figJVB}
\end{figure}

To obtain a quantitative description of the mixture vapor-liquid equilibria, one
state independent binary interaction parameter was adjusted to a single experimental data
point of either the vapor pressure or the Henry's law constant. Throughout, excellent
predictions were found at other state points, i.e., at other compositions or
temperatures as well as for the Henry's law constant, if it was adjusted to the vapor
pressure, or vice versa. Figures \ref{figJVA} and \ref{figJVB} show methane + methanol
as a typical example.

Furthermore, a set of 78 pure substances from prior work \cite{VSH01, SVH03b}
was taken to systematically describe all
267 binary mixtures of these components for which
relevant experimental VLE data are available. Again, per binary system, the single state
independent binary interaction parameter in the energy term was adjusted to only
one experimental value of the vapor pressure. The unlike energy parameter was thereby 
altered usually by less than 5 \%{} from the Berthelot rule. The mixture models
were validated regarding the vapor pressure at other state points as well as
the dew point composition, which is a fully predictive property in this context.
In almost all, i.e., 97 \%{} of the cases, the molecular models give excellent predictions
of the mixture properties. Compared to other works in the literature, this is by
far the largest investigation in this direction. It was facilitated by the
extensive computing equipment at the High Performance Computing Center Stuttgart. 

\begin{figure}[h!]
\centering
\includegraphics[width=0.85\textwidth]{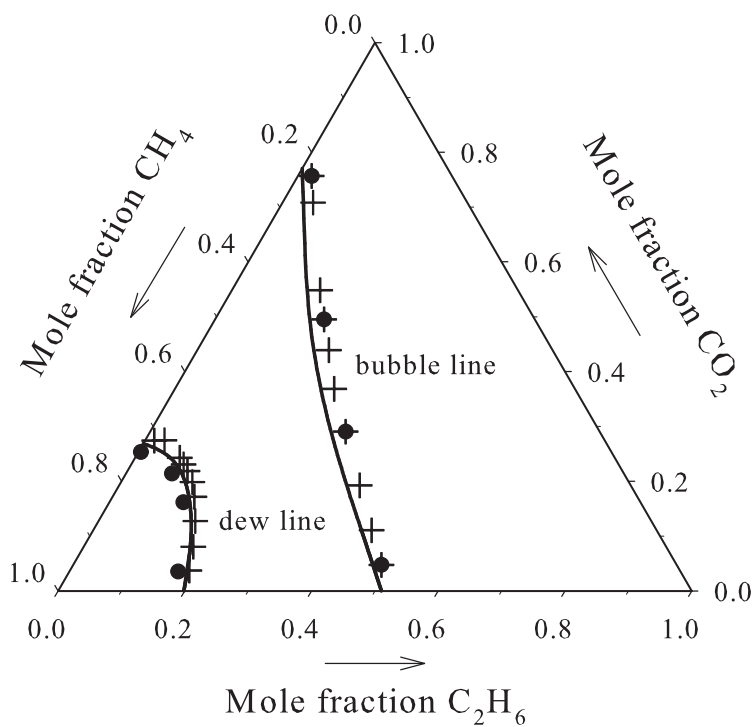}
\caption{
   Ternary vapor-liquid equilibrium phase diagram for
   the mixture \textsf{CH}$_4$ + \textsf{CO}$_2$ + \textsf{C}$_2$\textsf{H}$_6$ at 230 K and 4650 kPa:
   simulation data ($\bullet$), experimental data (+) of Knapp \etal{} \cite{KYZ90}, and
   Peng-Robinson equation of state (---).
}
\label{figternary}
\end{figure}

In the next step, all 33 ternary mixtures of these 78 components for which experimental 
VLE data is available were studied by molecular simulation. No adjustment to ternary
data was carried out at all so that the calculations were strictly predictive. By
comparing to experimental data, it was found that these predictions are very reliable
as there was practically always an excellent match. As an example, Fig.\ \ref{figternary} shows 
the ternary system consisting of methane, ethane, and carbon dioxide.
Again, the computational effort was substantial, publications in
the literature by other groups in this field typically cover one to two mixtures only.

\section{Vapor-liquid interface cluster criteria}

A suitable cluster criterion should achieve two goals: on the one hand,
it needs to distinguish the bulk liquid and the bulk vapor successfully in
every case -- even when the vapor is supersaturated, such as in the vicinity of a
droplet, or the liquid is undersaturated, such as in the vicinity of a bubble.
On the other hand, the cluster criterion should also
minimize noise fluctuations of the detected clusters to emphasize
the signal.

\begin{figure}[b!]
\centering
\includegraphics[width=0.85\textwidth]{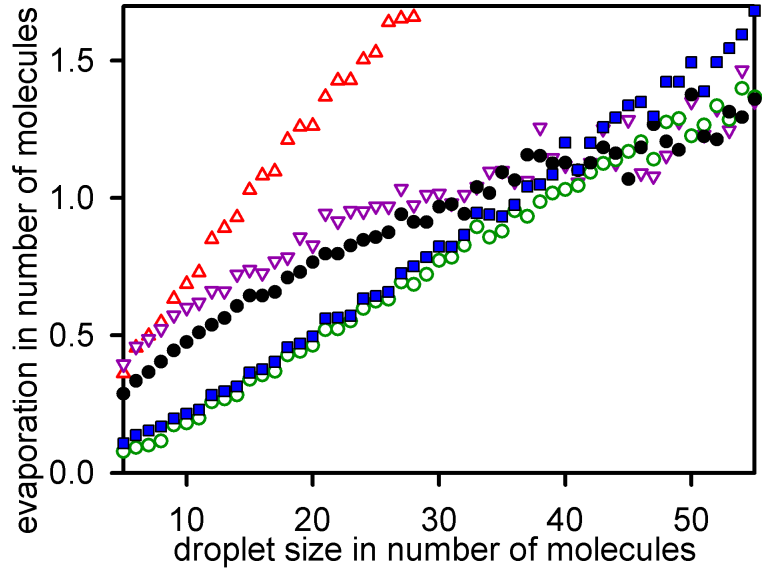}
\caption{
   Average number of molecules evaporating during a detection interval of 30 fs
   from droplets in a supersaturated vapor of carbon dioxide with $\temperature$
   = 237 K, $\density$ = 1.98 mol/l, and $\absnum$ = 1,050,000, determined according
   to various cluster criteria: Stillinger ($\circ$),
   ten Wolde and Frenkel with $\inta$ = 4 ($\bullet$),
   arithmetic mean with two ($\Delta$) and eight neighbors ($\nabla$) as well as
   geometric mean with two neighbors ($\square$).
}
\label{CO2evap237K}
\end{figure}

The following criteria for carbon dioxide were compared for this purpose using
a rigid two-center LJ plus point quadrupole (2CLJQ) model \cite{VSH01}:
\begin{itemize}
   \item{} Stillinger \cite{Stillinger63}: all molecules with a distance of 
           $\radiusS$ or less from each other are liquid and belong to the same
	   cluster (i.e., the same liquid phase or the same droplet). The
	   Stillinger radius was set to $\radiusS$ = 5.7 \AA{} for the present simulations.
   \item{} Ten Wolde and Frenkel \cite{RF98}: all molecules with at least $\inta$
           neighbors within a distance of $\radiusS$ belong to the liquid. They belong
	   to the same cluster as all other liquid molecules within a distance of $\radiusS$.
   \item{} Arithmetic mean, $\inta$ neighbors: a molecule is liquid if the density
           in the sphere containing its $\inta$ nearest neighbors exceeds
	   $(\liq{\density} + \vap{\density})/2$, where $\liq{\density}$ amd $\vap{\density}$
	   are the saturated liquid and vapor density, respectively.
	   The molecule belongs to the same cluster
	   as all other liquid molecules within the radius $\radius_\inta$,
	   which defines a sphere with the volume occupied by
	   $\inta + 1$ molecules at a density of $(\liq{\density} + \vap{\density})/2$.
   \item{} Geometric mean, $\inta$ neighbors: analogous, with a density threshold
           of $(\liq{\density}\vap{\density})^{1\slash{}2}$.
\end{itemize}
The simulations were conducted with the
massively parallel MD program \mardyn{} (the precursor implementation \textit{ls1 moldy} is
described by Bernreuther and Vrabec \cite{BV05}).
Figure \ref{CO2evap237K} shows that all of the discussed criteria are
applicable. The Stillinger criterion and the geometric mean density criterion
with two neighbors lead to the best results. It should be noted that
at high temperatures, i.e., near
the critical point, the Stillinger criterion becomes less reliable in distinguishing
the liquid from a supersaturated vapor than the geometric mean density criterion.

\section{Contact angle dependence on the fluid-wall interaction}

In cases where no
hydrogen bonds are formed between the wall and the fluid, vapor-solid and liquid-solid
interfacial properties mainly depend on the fluid-wall dispersive interaction,
even for hydrogen bonding fluids.
The truncated and shifted LJ (\LJTS) potential \cite{AT87}
\begin{equation}
   \potentialLJTS_{\moli\molj}(\distance_{\moli\molj})
      = \left\{ \begin{array}{l@{\quad \quad}l}
         4\LJenergy \left[\LJsize^{12}(\distance_{\moli\molj}^{-12} - \cutoff^{-12})
	    + \LJsize^{6}(\cutoff^{-6} - \distance_{\moli\molj}^{-6}) \right]
	   & \distance_{\moli\molj} < \cutoff \\
         0 & \distance_{\moli\molj} \geq \cutoff,
      \end{array} \right.
\end{equation}
with a cutoff radius of $\cutoff$ = 2.5 $\LJsize$
accurately reproduces the dispersive interaction
if adequate values for the size and energy parameters $\LJsize$ and $\LJenergy$
are speci\-fied. 
Due to the relatively small cutoff radius, all computations are accelerated,
while the descriptive power of the full LJ potential (without
a cutoff) is retained even for systems with phase boundaries \cite{VKFH06}.

For the present series of contact angle simulations, the \LJTS{} model --
with the potential parameters for methane, $\LJsize$ = 3.7241 \AA{}
and $\LJenergy\slash\kboltz$ = 175.06 K -- was
studied, extending a previous investigation
of the \vapour-liquid interface of the \LJTS{} fluid \cite{VKFH06}.
The \LJTS{} potential with the size and energy parameters $\LJsize' = \LJsize$
as well as
\begin{equation}
   \LJenergy' = \FWenergy\LJenergy,
\end{equation}
was also applied for the unlike interaction between the fluid molecules and the
wall atoms, with the same cutoff radius as for the fluid.
The wall was modeled as a system of coupled harmonic oscillators with different
spring constants for transverse and longitudinal motion,
adjusted to simulation results for graphite that were
obtained with a rescaled version of
the Tersoff \cite{Tersoff88} potential.

\begin{figure}[h]
\centering
\includegraphics[width=8.8cm]{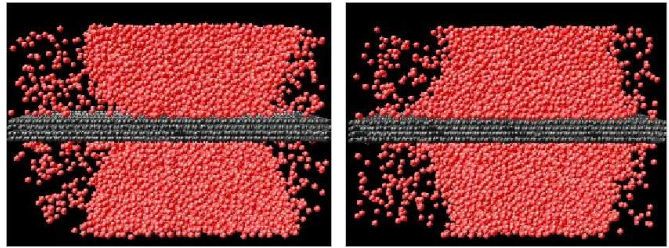}
\caption{Simulation snapshots for a reduced
fluid-wall dispersive energy $\FWenergy$ of 0.07 (left)
and 0.13 (right) at a temperature of 0.88 $\FWenergy\slash\kboltz$.}
\label{fig04}
\end{figure}

The simulations were carried out with the \mardyn{} program. \Vapour{} and
liquid were independently equilibrated in homogeneous simulations for 10 ps. This was followed by
200 ps of equilibration for the combined system, i.e., a liquid meniscus surrounded
by \vapour{}, with a graphite wall consisting of four to seven
layers, cf. Fig.\ \ref{fig04}.
A periodic boundary condition was applied to the system, leaving a channel
with a diameter of 27 $\LJsize$ between the wall and its periodic image.
The contact angle was determined from the density profiles by averaging
over at least 800 ps after equilibration.

Liquid menisci between graphite walls were simulated 
for a reduced fluid-wall dispersive energy $\FWenergy$ between 0.07 and 0.16
at temperatures of 0.73, 0.88, and 1 $\LJenergy\slash\kboltz$.
Note that the triple point temperature of the \LJTS{} fluid
is about 0.65 $\LJenergy\slash\kboltz$ \cite{MPSF08}
while the critical temperature is $\temperaturecrit$ =
1.0779 $\LJenergy\slash\kboltz$ \cite{VKFH06},
so that the entire regime of stable \vapour-liquid coexistence was
covered.

\begin{figure}[h]
\centering
\includegraphics[width=8.8cm]{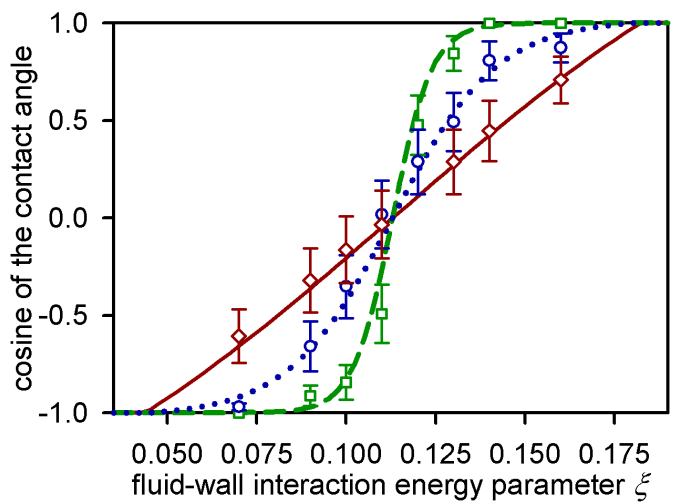}
\caption{Simulation results and correlation for the contact angle
in dependence of the reduced fluid-wall dispersive energy $\FWenergy$
at temperatures of 0.73 (diamonds and solid line), 0.88 (circles and
dotted line) as well as 1 (squares and dashed line) $\LJenergy\slash\kboltz$.}
\label{fig13}
\end{figure}

High values of $\FWenergy$ correspond to a strong attraction between fluid
and wall compoents, leading to a contact angle $\contactangle$ smaller than $\degree{90}$, 
i.e., to partial ($\contactangle > \degree{0}$) or full ($\contactangle = \degree{0}$)
wetting of the surface.
For a higher fluid-wall dispersive energy, the extent of
wetting increases, cf.\ Fig.\ \ref{fig04}.
The transition
from obtuse to acute contact angles, i.e.,
   $\contactangle(\temperature, \cpenergy) = \degree{90},$
occurs at a temperature independent value $\cpenergy \approx 0.113$ of
the fluid-wall dispersive energy, as can be seen in Fig.\ \ref{fig13}.
Furthermore, the symmetry law
\begin{equation}
   \cos\contactangle(\temperature, \cpenergy - \Delta\FWenergy)
      = - \cos\contactangle(\temperature, \cpenergy + \Delta\FWenergy),
\label{eqn:sym}
\end{equation}
is valid over the whole relevant range of temperatures and magnitudes of
the fluid-wall dispersive energy.
Figure \ref{fig13} also shows that there is a narrow
range of $\FWenergy$ values that lead to the
formation of a contact angle, as opposed to total dewetting or wetting.
As the temperature decreases and the \vapour-liquid
surface tension $\tensionvl(\temperature)$ increases, the
contact angle approaches $\degree{90}$.

\section{A sterically accurate generic benzyl alcohol model}

Benzyl alcohol ($\textsf{C}_{6}\textsf{H}_{5}$--$\textsf{CH}_{2}\textsf{OH}$)
is widely used as a solvent for paints and inks.
However, it is classified as a harmful
substance (Xn) and should not be inhaled, nor used at high temperatures where
it exhibits a high vapor pressure. 

\begin{figure}[b!]
\centering
\includegraphics[width=0.85\textwidth]{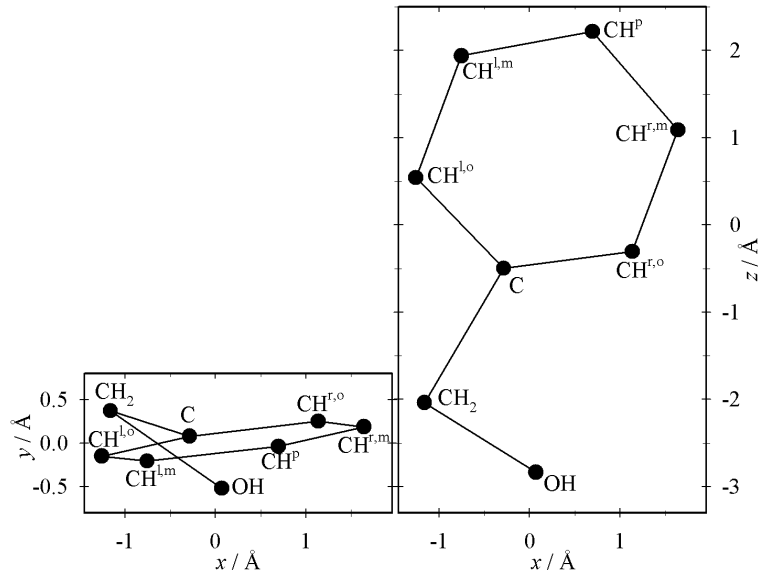}
\caption{
   Coordinates of the LJ sites for the present benzyl alcohol model.
}
\label{merker2}
\end{figure}

The basis for a new rigid molecular model of benzyl alcohol was
determined by \textit{ab initio} calculations with the GAMESS (US) quantum
chemistry package \cite{SBBEGJKMNSWDM93}, obtaining
the quadrupole moment as well as the equilibrium positions of the nuclei as
illustrated in Fig.\ \ref{merker2}.

Further potential parameters were taken from accurate empirical molecular models for
related fluids, leading to a sterically accurate generic model, cf.\ Tab. \ref{tab_coord},
that can be used as a starting point for parameter optimization. In particular,
point charges as well as $\LJsig$ and $\LJeps$ parameters for the hydroxyl group
were taken from the ethanol model of Schnabel \textit{et al.} \cite{SVH05}. Moreover,
the $\LJsig$ and $\LJeps$ values of the corresponding LJ interaction site of the
Merker \etal{} \cite{MGVH09} cyclohexanol model were used for the \textsf{CH}$_2$ group,
while the LJ parameters for the \textsf{CH} and \textsf{C} centers were set according
to the Huang \textit{et al.} \cite{HHV10} models of benzene and phosgene, respectively.
\begin{table}[h!]
\noindent
\caption{
   Coordinates and parameters of the LJ sites, and the point charges, and the point quadrupole
   for the present benzyl alcohol model. Bold characters indicate represented atoms.
}
\label{tab_coord}
\medskip
\begin{center}
\begin{tabular}{lccccccc} \hline\hline
Interaction & $x$   & $y$   & $z$   & $\LJsig$ & $\LJeps/k_\mathrm{B}$ & $q$ & $Q$ \\
site        & \AA{} & \AA{} & \AA{} & \AA{}    & K                     & $e$ & $D$\AA{} \\ \hline
\textbf{C}H$^\textrm{p}$ & \phantom{-}0.695 & -0.037 & \phantom{-}2.218 & 3.247 & \phantom{0}88.97&& \\
\textbf{C}H$^\textrm{l,m}$ & -0.753 & -0.204 & \phantom{-}1.941 & 3.247 & \phantom{0}88.97&& \\
\textbf{C}H$^\textrm{r,m}$ & \phantom{-}1.638 & \phantom{-}0.189 & \phantom{-}1.090 & 3.247 & \phantom{0}88.97&& \\
\textbf{C}H$^\textrm{l,o}$ & -1.255 & -0.149 & \phantom{-}0.543 & 3.247 & \phantom{0}88.97&& \\
\textbf{C}H$^\textrm{r,o}$ & \phantom{-}1.134 & \phantom{-}0.250 & -0.304 & 3.247 & \phantom{0}88.97&& \\
\textbf{C} & -0.285 & \phantom{-}0.080 & -0.497 & 2.810 & \phantom{0}10.64&&  \\
\textbf{C}H$_2$ & -1.160 & \phantom{-}0.372 & -2.039 & 3.412 & 102.2\phantom{0} & +0.2556& \\
\textbf{O}H & \phantom{-}0.070 & -0.516 & -2.835 & 3.150 & \phantom{0}85.05 & -0.6971& \\
O\textbf{H} & \phantom{-}0.191 & -1.430 & -2.614 & & & +0.4415& \\
Benzyl & \phantom{-}0\phantom{.000} & \phantom{-}0\phantom{.000} & \phantom{-}0\phantom{.000}
       & & & & 2.534 \\ \hline\hline
\end{tabular}
\end{center}
\end{table}

\begin{figure}[t!]
\centering
\includegraphics[width=0.85\textwidth]{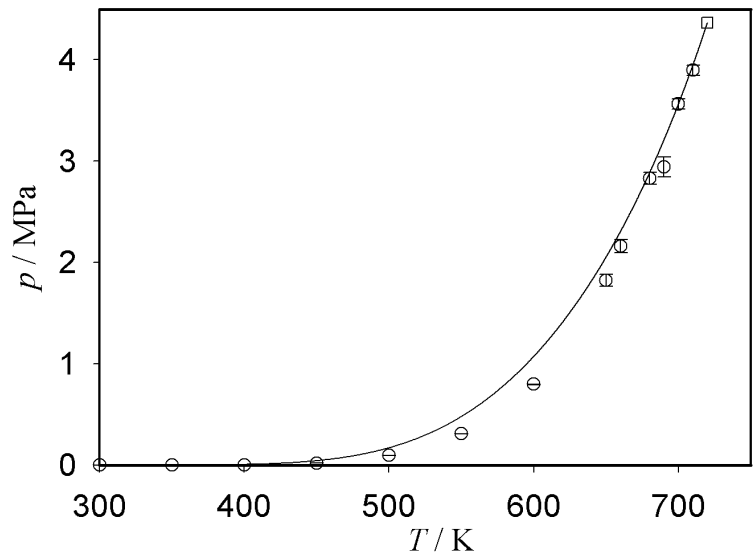}
\caption{
   Vapor pressure of benzyl alcohol according to a correlation \cite{RWOYZDD06} based on
   experiments (---) as well as present molecular
   simulation data ($\circ$).
}
\label{merker}
\end{figure}

VLE properties of the generic model were calculated using the \mstwo{} program,
leading to an overall satisfactory first approximation, considering that all
a posteriori adjustments were absent.
The vapor pressure is more accurate at high temperatures, cf.\ Fig.\ \ref{merker}.

\section{Computing performance}

The scalability of the \mardyn{} program was measured on the \textit{cacau}
supercomputer for simulation scenarios involving methane, represented by the
\LJTS{} fluid, as well as graphite, modeled by a rescaled version of
the Tersoff \cite{Tersoff88} potential.
MPI parallelization was applied according to a spatial domain decomposition scheme with
equally sized cuboid subdomains and a cartesian topology based on linked cells \cite{BV05}.

\begin{figure}[h!]
\centering
\includegraphics[width=0.85\textwidth]{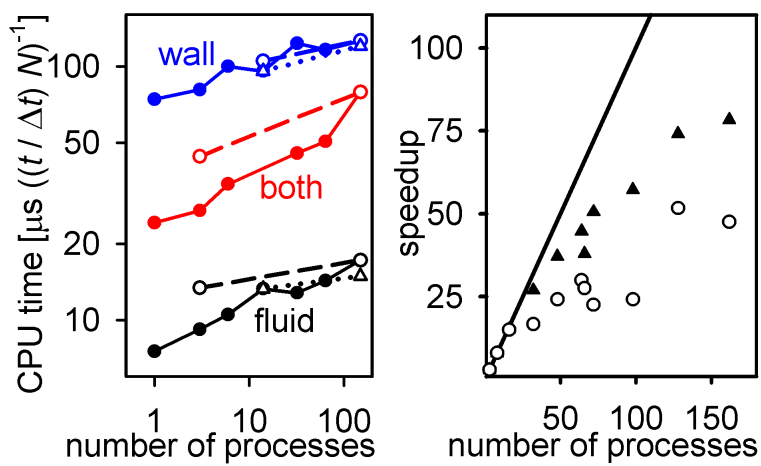}
\caption{Left: Total CPU time, i.e., execution time multiplied with the
number of parallel processes, per time step and interaction site for weak scaling
with 3,000 (dashed lines / circles) and 32,000 (dotted lines / triangles)
interaction sites per process as well as strong scaling with 450,000
interaction sites (solid lines / bullets), using isotropic spatial domain
decomposition.
Right: speedup for a system of liquid methane between graphite walls
with 650,000 interaction sites, where isotropic (circles) and channel geometry
based (triangles) spatial domain decomposition was used;
the solid line represents linear speedup.}
\label{0RAB}
\end{figure}

Often the best solution is an isotropic decomposition that minimizes
the surface to volume ratio of the spatial subdomains.
For the simulation of homogeneous systems, this approach is quite efficient.
That is underlined by the weak and strong scaling behavior of \mardyn{} for typical
configurations, shown in Fig.\ \ref{0RAB} (left), in cases where supercritical
methane (\qq{fluid}) at a density of 10 mol/l and solid graphite (\qq{wall}) were considered
with a system size of up to 4,800,000 interaction sites, representing the
same number of carbon atoms and methane molecules here. Graphite
simulations, containing only carbon atoms, scale particularly
well, due to a favorable relation of the delay produced by communication
between processes to the concurrent parts, 
i.e., the actual intermolecular interaction computation,
which is much more expensive for the Tersoff potential than the \LJTS{}
potential.

The simulation of combined systems,
containing both fluid and solid interaction sites,
is better handled by a channel geometry based decomposition scheme, where
an approximately equal portion of the wall and a part of the fluid
is assigned to each process, cf.\ Fig.\ \ref{0RAB} (right).
In the general case, where spatial non-uniformities do not match any cartesian grid,
a flexible topology has to be used. An approach based on
$k$-dimensional trees \cite{Bentley75},
implemented in a version of \mardyn{}, showed clearly improved results with
respect to the scaling of inhomogeneous systems.

\section{Conclusion}

The intermolecular interaction of small hydrogen bonding molecules like mono-
and dimethylamine can be described by simple LJ based united-atom molecular models
with point charges. Such computationally efficient models were
applied to binary mixtures with non-hydrogen bonding components regarding VLE
properties. Accurate predictions, covering a broad range of temperatures and
compositions, were obtained, regardless whether the state independent binary
interaction parameter was adjusted to Henry's law constant or vapor pressure.

For a two-center LJ plus quadrupole model of carbon dioxide,
a comparison of cluster criteria with the purpose of accurately
detecting the vapor-liquid
phase boundary gave overall support to the geometric mean density criterion applied
to the sphere consisting of a molecule and its two nearest neighbours.
The Stillinger criterion was found to be particularly adequate at low temperatures.
For the \LJTS{} fluid, the contact angle formed between the vapor-liquid interface and
a wall was determined by canonical ensemble MD simulation while the mangitude
of the dispersive fluid-wall interaction was varied. Over the whole
temperature range between triple point and critical point, the contact angle
dependence on the fluid-wall dispersive energy obeys a simple symmetry law.

A sterically realistic model for benzyl alcohol was presented, showing within the framework
of LJ sites with point polarities and electric point charges
that the generic molecular modeling approach can lead
to a good starting point for parameter optimization with respect to
VLE properties such as the vapor pressure.

The scalability of the \mardyn{} program
was assessed and found to be acceptable.
MD simulations of methane confined between graphite
walls with up to 4,800,000 interaction sites, i.e., carbon atoms and
methane molecules, were conducted to demonstrate
the viability of the program.

The authors would like to thank Martin Bernreuther, Martin Buchholz, Domenic Jenz,
and Christoph Niethammer for contributing to the \mardyn{} code, Stephan Deublein,
Bernhard Eckl, Gimmy Fern{\'a}ndez Ram{\'i}rez, Gabriela Guevara Carri{\'o}n,
and Sergei Lishchuk for contributing to the \mstwo{} code,
Ioannis Pitropakis for performing simulation runs of binary mixtures,
as well as Calin Dan, Franz G\"ahler, Sebastian Grottel, 
Jens Harting, Martin Hecht, Ralf Kible, and Guido Reina for frank discussions
and their persistent support.


\end{document}